\documentclass[twocolumn,amsmath,amssymb,floatfix]{revtex4-1}

\def\Tr{\textrm{Tr~}}
\def\half{\frac{1}{2}}
\def\lb({\left(}

\def\hbar{\hspace{0pt}\raisebox{1pt}{$-$} \hspace{-7pt} h}

\newcommand{\eq}[1]{\begin{equation} #1 \end{equation}}
\newcommand{\eqs}[1]{\begin{eqnarray} #1 \end{eqnarray}}
\def\a{\alpha}

\def\g{\gamma}
\def\d{\delta}
\def\e{\epsilon}
\def\z{\zeta}
\def\ee{\eta}

\def\l{\lambda}
\def\m{\mu}
\def\n{\nu}

\def\p{\pi}
\def\r{\rho}
\def\s{\sigma}

\def\U{\Upsilon}

\def\oo{\omega}

\def\G{\Gamma}
\def\D{\Delta}

\def\p{\pi}
\def\r{\rho}
\def\S{\Sigma}

\def\U{\Upsilon}

\usepackage{amssymb, amsmath, mathrsfs,verbatim, subfig,bbm}
\usepackage[dvips]{graphicx}
\begin{document}
\title{Generalised Holographic Electroweak Symmetry Breaking Models and the
Possibility of Negative $\hat{S}$}
\author{Mark Round}
\date{\today}
\affiliation{
Swansea University, School of Physical Sciences, \\
Singleton Park, Swansea, Wales, UK}
\begin{abstract}
Within an AdS/CFT inspired model of electroweak symmetry breaking the
effects of
various boundary terms and modifications to the background are studied. 
The effect
on the $\hat{S}$ precision parameter is discussed with particular
attention to its
sign and whether the theory is unitary when $\hat{S}<0$.  Connections between the various possible AdS slice models of symmetry breaking are discussed.
\end{abstract}
\maketitle
\section{Introduction}

Models of electroweak symmetry breaking are compared to experiment through
the precision electroweak parameters~\cite{Peskin:1990zt, Peskin:1991sw,
Barbieri:2004qk}.  The parameter of primary interest in
Higgsless~\cite{Csaki:2003zu, Cacciapaglia:2004rb,Cacciapaglia:2004jz} and
technicolor models~\cite{Weinberg:1975gm, Susskind:1978ms, Weinberg:1979bn} is $\hat{S}$.  Much effort has gone into understanding
how one can produce models where $\hat{S}$ can be tuned within experimentally
acceptable bounds.  Recently studies pursued this end using 5D
models formulated on a slice of a space that asymptotes towards AdS in the
UV~\cite{Csaki:2003zu, Cacciapaglia:2004rb, Cacciapaglia:2004jz, DaRold:2005zs, Erlich:2005qh,
Hirn:2006wg, Piai:2006hy, Barbieri:2003pr, Agashe:2007mc, Dietrich:2005jn,
Carone:2006wj, Hong:2006si, Fabbrichesi:2008ga, Piai:2007ys}.

In developing holographic models that produce an arbitrarily small
$\hat{S}$ parameter it is interesting to ask whether $\hat{S}$ can be made
negative.  It is known that the difference in vector and axial two-point
functions is strictly positive~\cite{Witten:1983ut}, suggesting that
$\hat{S}$ is too (in a physically acceptable scenario).  In addition, a
 unitary four-dimensional model with negative $\hat{S}$ has never
been found; except for models where new Majorana fermions are
added into the Standard Model~\cite{Dobrescu:1997kt}.  This can lead to
$\hat{S}<0$ but one is using matter fields to dial $\hat{S}$ rather than purely the mechanism of electroweak symmetry breaking.  We will not be interested in the effects of matter on $\hat{S}$.  Ignoring this possibility, the absence of a negative $\hat{S}$ model and the result of Witten~\cite{Witten:1983ut} suggest that it might not be not possible to make $\hat{S}<0$ without introducing pathologies.

Previous literature has made progress in addressing the possibility of
$\hat{S}<0$ within holographic models.  Early work showed that in the most
basic set-ups of a pure AdS geometry and no boundary terms
$\hat{S}>0$~\cite{Hong:2006si}.  Meanwhile work in holographic models of QCD led to
the idea of symmetry breaking by the geometry itself -- by using  different
geometries for the axial and vector fields~\cite{Hirn:2005vk}.  This idea
was applied to technicolor and a numerical computation of the $\hat{S}$
parameter showed regions where $\hat{S}<0$~\cite{Hirn:2006nt}.  The
Lagrangian considered contains a spurion and as such encompasses a class
of models.  Although this work shows it is possible to construct a model
where $\hat{S}<0$, it is not apparent that this can be done within the
context of a consistent effective theory.  Further development of these
ideas was made by studying a realistic example. The authors
of~\cite{Hirn:2006wg} studied an example model and
discussed how the result of a negative $\hat{S}$  could be understood
in 4D.  The effects on $\hat{S}$ by adding more general terms into the
Lagrangian was studied by using St\" uckelberg
fields~\cite{Dietrich:2008up}.  Other authors, with particular attention
to the sign of $\hat{S}$, have discussed symmetry breaking by boundary
terms and a numerical scan of possible VEV profiles~\cite{Agashe:2007mc}.

In this paper  we will focus on the Goldstone modes and whether or not
they are negative-norm states.  This will be done by computing the sign of
the Goldstone kinetic term.  This exercise will be carried out with the inclusion of
all extra bulk and boundary terms (relevant to the discussion of the sign
of $\hat{S}$)  that can be added to a basic holographic model of
electroweak symmetry breaking.  As this introduction has shown, some parts
of this study already exist in the literature.  We will include all the
(leading-order) terms possible and discuss their interplay in a systematic
and analytic way.

\section{Five Dimensional $SU(2)\times SU(2)$ Action\label{modbuild}}
This work will study an $SU(2)_L \times SU(2)_R$ gauge symmetry broken to 
the diagonal  $SU(2)$ by a bifundamental scalar field, $H$, in five
dimensions.  The geometry  is defined by the metric
\eq{\label{metric}
ds^2 =g_{MN}dx^Mdx^N = \oo^2(y) \left( \ee_{\m\n}dx^\m dx^\n - dy^2\right),
}with $\ee_{\m\n}$ the Minkowski metric of signature $(+,-,-,-)$.  Here,
$y\in [L_0,L_1]$ is the $5^\textrm{th}$ dimension.  The scales $L_0$ and
$L_1$ represent the UV and IR cut-offs of the theory respectively. 
Lorentz indices running over the 5D manifold will be denoted by uppercase
Latin indices.  Lowercase Greek indices run over the first four of the
five co-ordinates.  Requiring  that the space asymptotes towards AdS gives
the constraint
\eq{\label{metriclimit}
\oo(y) \rightarrow \frac{L}{y}  \textrm{ as } y\rightarrow L_0 .
}

If the gauge fields are denoted by $L^a_M$ and $R^a_M$, then the matrix-valued fields $L_M = L^a_M \s^a/2 $ and $R_M = R^a_M \s^a/2 $ (for $\s^a$,
$a = 1,2,3$,  the Pauli matrices) can be used to write out the possible
Lorentz and gauge invariant operators.  The relevant operators are of the
form
\eq{\label{treeops}
m^2|H|^2  , \,\Tr |D_M H|^2, \,\Tr L_{MN}L^{MN}, \, \Tr R_{MN}R^{MN}.
}The field strength and covariant derivative are
\eqs{
L_{MN} &=& \partial_M L_N - \partial_N L_M -i g_L [L_M, L_N],\\
D_M H &=& \partial_M H -i g_L L_M H + i g_R H R_M,
}with $g_L$($g_R$)  the gauge coupling of the $L$($R$) field.  Localised
on the boundaries there are further relevant operators,
\eq{
\Tr |D_\m H|^2, \, m'^2 \Tr |H|^2, \, \Tr L_{\m\n}L^{\m\n} , \, \Tr
R_{\m\n}R^{\m\n}
}The operator $HH^\dagger H H^\dagger$ is marginal on the boundaries.

Gauge kinetic operators $\Tr L_{\m\n}L^{\m\n} $ and $\Tr R_{\m\n} R^{\m\n}$
when localised on the UV  boundary act as counterterms to the UV boundary
theory.  The co-efficients of the terms can be dialed so as to remove
spurious dependence upon the UV cut-off $1/L_0$.  This is the process of
holographic renormalisation.  See ~\cite{Round:2010kj, Skenderis:2002wp}
for a discussion of holographic renormalisation in this context.

In addition to the operators already discussed there might be 
higher-order corrections.  The first corrections to the bulk action are,
\eq{\label{hightreeops}
\Tr |H^\dagger D_M H|^2 , \, \Tr L_{MN}H R^{MN} H^\dagger
}which are the 5D analogues of the operators corresponding to the
$\hat{T}$ and $\hat{S}$ electroweak precision parameters
respectively~\cite{Peskin:1990zt,Peskin:1991sw, Barbieri:2004qk}.  It is
the $\hat{S}$ operator that will be of primary interest in this work. 
More formally, one could argue that $\hat{T}$ can be suppressed  by
invoking a custodial symmetry.  In this work there is a custodial symmetry which allows us to neglect the $\hat{T}$-like operator.

In addition to the operators listed in eqs. (\ref{treeops}) and
(\ref{hightreeops}) one can write potentials for both the scalar fields,
the Einstein-Hilbert term, couplings of the vector and scalar field to
gravity and topological terms.  In this work we will \textit{not}
back-react the fields on the geometry in neglecting backreaction we work
under the probe approximation.  Interactions of the gauge fields are
suppressed by the large-$N_c$ limit ($N_c$ the degree of gauge group in
the dual theory) which implies that $g^2_{L(R)} \sim  1/N_c$ and therefore
in this work the 3-point and 4-point boson vertices will be set to zero.  
For a discussion of the large-$N_c$ limit see ~\cite{Round:2010kj}.

In summary the leading order bulk action for this model is
\eqs{\label{action}
S_0 &=& \int \sqrt{g} d^4x dy \Tr\bigg[-\half
g_{MN}g_{PQ}[L^{MP}L^{NQ}+R^{MP}R^{NQ} ]\nonumber\\&+&g_{MN} (D^M
H)(D^NH)^\dagger +m^2  |H|^2 \bigg]
}where the metric is defined in eq.~(\ref{metric}).  This is the simplest
construction possible for this model.  Additional terms will be discussed
in later sections.  

To break the electroweak symmetry the scalar field
obtains a VEV
\eq{\label{pidef}
\langle H \rangle = \half  f(y) e^{i\p (x^\m ,y)}\mathbbm{1},
}where $\p =\p^a \s^a/2$ are the pion fields associated to the breaking
$SU(2)\times SU(2)\rightarrow SU(2)$.  $f(y)$ is a VEV profile and
$\mathbbm{1}$ is the unit matrix.  Notice that the exponent does not
include a factor of $f(y)$ for convenience.

It is useful to perform calculations in the basis of axial (A) and vector
(V) fields defined as
\eqs{
A^M &=& \frac{g_L L^M - g_R R^M}{\sqrt{g_L^2+g_R^2}},\\
V^M &=&\frac{g_R L^M + g_L R^M}{\sqrt{g_L^2+g_R^2}}.
}

The bulk equations of motion for the gauge fields can be written by first
Fourier transforming in 4D and separating the gauge field out into a
$y$-dependent function and a $q^2$ dependent function,
\eqs{
A^\m (x^\m,y)& \rightarrow & a(q^2,y)\hat{A}^\m (q),\\
V^\m (x^\m,y)& \rightarrow & v(q^2,y)\hat{V}^\m (q).
}Using this notation the equations of motion are,
\eqs{\label{Veqnmotion}
\left[\frac{1}{\oo}\partial_y \oo\partial_y +q^2\right]\!
v(q^2,y)\!&=& \! 0\\\label{Aeqnmotion}
\left[\frac{1}{\oo}\partial_y \oo\partial_y +\! q^2 \! -
\!\frac{1}{4}(g_L^2+g_R^2)\oo^2f^2(y)\right]\! a(q^2,y)\! &=& \!0
}The IR boundary conditions, found by demanding that the field variations
of  the action eq. (\ref{action}) vanish on the IR boundary, are
\eq{
\left. \begin{array}{c}
-\oo(y) \partial_y v(q^2,y)\vert_{y=L_1}\\
-\oo(y) \partial_y a(q^2y)\vert_{y=L_1}
\end{array}\right\}
 =\textrm{IR boundary terms}
}For the case where there are no IR boundary terms in the action the expressions above both
vanish.  Substituting the solutions of the equations of motion back into
the action leaves the 4D UV boundary theory with action
\eqs{
S &=&\int \hat{A}^\m(-q) \Pi_A(q^2) P_{\m\n} \hat{A}^\n(q) \nonumber\\
&+& \hat{V}^\m(-q) \Pi_V(q^2) P_{\m\n} \hat{V}^\n(q) d^4q,\\
\Pi_V(q^2) &=& \left.\oo(y) \frac{\partial_y
v(q^2,y)}{v(y)}\right\vert_{y=L_0} + \textrm{UV terms},\label{Vvacpol}\\
\Pi_A(q^2) &=& \left.\oo(y) \frac{\partial_y
a(q^2,y)}{a(y)}\right\vert_{y=L_0}+ \textrm{UV terms}\label{Avacpol}.
}where $P_{\m\n}=\ee_{\m\n}-q^\m q^\n/q^2$ and $\Pi_{A,V}$ are vacuum
polarisations.  The expression `UV terms' refers to contributions to $\Pi_{A,V}$ that originate from operators in the action that are localised on the UV boundary.

In the small-$L_0$ limit, there is a logarithmic divergence associated to
$\Pi_{V,A}$.  This can be cured by adding a boundary counterterm of the
form
\eqs{
S_{C.T.} &=& \int \sqrt{g}d^4xdy\d(y-L_0)Z g_{\m\n}g_{\r\s}\nonumber\\
&\times& \left[-\half \Tr L^{\m\r}L^{\n\s}
-\frac{1}{4}R^{3\m\r}R^{3\n\s}\right]
}to the action ($S=S_0+S_{C.T}$) and adjusting $Z$ to remove dependence
upon the cut-off $L_0$.  Choosing to renormalise only $R^3$ means that the
associated charged states decouple.  In this way the lightest modes of the
spectrum are that of the standard model.  Above these modes a full copy of
the broken $SU(2)\times SU(2)$ is realised at each level in the tower of
states.  For further discussion, including the gauge invariance of
$S_{C.T.}$ see \cite{Round:2010kj}.

\subsection{Precision Electroweak Parameters}
The precision parameters $\hat{S}$ and $W$ are defined as
\eqs{
\hat{S} &=& \frac{g}{g'}\left. \frac{d}{dq^2} \Pi_{W^3
B}(q^2)\right\vert_{q^2=0}\\
W &=&\half m_W^2 \left. \frac{d^2}{d(q^2)^2}\Pi_{W^3W^3}\right\vert_{q^2=0}
}where $\Pi_{W^3B}$ and $\Pi_{W^3,W^3}$ are elements in the vacuum
polarisations expressed using the $(W^3,B)$ basis.  The couplings $g$ and
$g'$ are the $SU(2)$ and $U(1)$ couplings in the standard model.  The
vacuum polarisations are normalised so that
\eqs{
\left.\frac{d}{dq^2}\Pi_{BB}(q^2)\right\vert_{q^2=0}&=&
\left.\frac{d}{dq^2}\Pi_{+-}(q^2)\right\vert_{q^2=0}=1 \\
 \Pi_{+-}(0)&=&m_W^2
}where $\pm$ refers to the $W^\pm$ fields and $m_W$ is the $W^\pm$ mass. 
$B$ is the $U(1)$ gauge field of the Standard Model.
\subsection{Elementary Example\label{elemcase}}
In this section the simplest construction of an AdS model of electroweak
symmetry breaking is studied.  This will provide a basis for further
discussion, define some notation and remind the reader of material
available in the literature~\cite{Maldacena:1997re, Aharony:1999ti,
Witten:1998qj, Gubser:1998bc, Randall:1999ee, Csaki:2003zu,
Cacciapaglia:2004rb, Cacciapaglia:2004jz, DaRold:2005zs, Erlich:2005qh,
Hirn:2006wg, Piai:2006hy, Barbieri:2003pr, Agashe:2007mc, Dietrich:2005jn,
Carone:2006wj, Hong:2006si, Fabbrichesi:2008ga, Piai:2007ys}.

By using the effective action eq.(\ref{action}) and a given background
$\oo$ the VEV profile $f(y)$ can be found by the solving the equations of
motion for $H$.  The simplest case is an AdS geometry which implies the VEV profile
that solves the equations is schematically $f(y) \sim c_1 y^\D + c_2 y^{4-\D}$.  Now pick just one of the two power-laws, our choice is that
\eq{\label{simpleVEV}
f(y) = \U\left(\frac{y}{L_1}\right)^\D \textrm{ and } \oo(y) = \frac{L}{y},
}$\U$ is the VEV and  $\D$ is the anomalous dimension of the condensate. 
In this case the 5D mass of $H$ is related to the VEV profile through
\eq{
m^2 =\frac{\D(\D-4)}{L^2}.
}The early sections of this paper will use eq. (\ref{simpleVEV}) as the
VEV profile.  In later sections contributions will be added to the action
that imply a different VEV profile.

The computation of $\hat{S}$ proceeds by solving the equations of motion
for the profiles $v(q^2,y)$ and $a(q^2,y)$, substituting into the vacuum
polarisations given in Eqs. (\ref{Vvacpol}) \& (\ref{Avacpol}) and taking the small-$L_0$ limit.  The counterterm $Z$ is
dialed to remove the divergences that occur.  The renormalised 
$\Pi_{V,A}$ can be rotated back into the $(W^3,B)$ basis and $\hat{S}$
extracted.

In general the equations of motion do not have closed form solutions. 
However, for the case of eq. (\ref{simpleVEV}), the vector equation has a
closed form solution.  Solving eq. (\ref{Veqnmotion}) with the choice of
eq. (\ref{simpleVEV}) and expanding the resulting expression for
$\Pi_V(q^2)$ in powers of $L_0$ one encounters a divergence that is cured
by the counterterm $Z$ if
\eq{
Z= L_0\log \frac{L_0}{L_1}  + \frac{L_0}{\e^2}.
}In this case $\Pi_V$ is  rendered free from spurious dependence on the UV
cut-off $L_0$ but one  introduces a new parameter $\e$.  With the vacuum
polarisation free from divergences the limit $L_0 \rightarrow 0 $ can be
taken.

To solve the axial equation of motion define~\cite{Hong:2006si}
\eq{
P(q^2,y) = \oo(y)\partial_y \log a(q^2,y)
}which satisfies a differential equation derived from eq. (\ref{Aeqnmotion}),
\eq{
\frac{1}{\oo}P'(q^2,y) +\frac{1}{\oo^2}P^2(q^2,y) +q^2
-\frac{1}{4}(g_L^2+g_R^2)\oo^2 f^2(y)=0.
}Expand $P$ in powers of $q^2$ so that $P(q^2,y)=P_0(y)+q^2P_1(y)+\ldots$
which satisfy the differential equations
\eqs{
\frac{1}{\oo}P_0' +\frac{1}{\oo^2}P_0^2
-\frac{1}{4}(g_L^2+g_R^2)\oo^2f^2&=&0,\\
\frac{1}{\oo}P_1' +2\frac{1}{\oo^2}P_0P_1 +1&=&0.
}For the case of eq. (\ref{simpleVEV}), where the boundary condition is
written as $P(q^2,L_1)=0$, both equations have closed form solutions.  The
expressions that solve the equations of motion can be  found in Appendix
A.

A simple expression for $\hat{S}$ can obtained in the same regime of
$(g_L^2+g_R^2)L^2 \U^2 \ll 1$ and $\D>1$
\eq{
\hat{S} \simeq g_L^2 \frac{1}{16}\e^2 \frac{\D+1}{\D^2}\U^2 L^2\simeq
\half \frac{\D^2-1}{\D^2}m_W^2 L_1^2.
}where the $W$ mass $m_W^2 = g_L^2 m_Z^2/(g_L^2+g_R^2)$ has been
introduced.  Further details of this calculation can be found in the
literature~\cite{Maldacena:1997re, Aharony:1999ti, Witten:1998qj,
Gubser:1998bc, Randall:1999ee, Csaki:2003zu, Cacciapaglia:2004rb,
Cacciapaglia:2004jz, DaRold:2005zs, Erlich:2005qh, Hirn:2006wg,
Piai:2006hy, Barbieri:2003pr, Agashe:2007mc, Dietrich:2005jn,
Carone:2006wj, Hong:2006si, Fabbrichesi:2008ga, Piai:2007ys, Piai:2006hy}.  For this work the conclusion to draw is that
$\hat{S}$ is positive definite in the most simple case we study, as
already noted in \cite{Hong:2006si}.

Now consider the same background and VEV profile but with a modified
action that no longer contains gauge fields.  The $SU(2)\times SU(2)$
symmetry is now a global symmetry of the scalar field $H$.  In doing this
a theory describing the scalar field $\p$ is obtained.  Recall that $\p$ was defined in eq. (\ref{pidef}) as the scalar rotations about the VEV.  Working in the unitary gauge of the full local theory, the pions are Higgsed away: only when the gauge couplings are turned off are the pions manifest (and in a unitary gauge).  This allows one to
ensure that the proposed action for some model of electroweak symmetry
breaking satisfies the conditions of a sensible theory -- in this case
that the theory is free from negative-norm states.
\eqs{
S &=& \int\sqrt{g} d^4x dy g_{MN}\Tr (\partial^M H)(\partial^N H)^\dagger
+m^2 \Tr |H|^2 \nonumber\\
&=& \sum_a \frac{1}{8}\int d^4xdy\p^a [-\oo^3(y)f^2(y) \ee_{\m\n}
\partial^\m  \partial^\n \nonumber\\
&+&  \partial_y
\oo^3(y)f^2(y)\partial_y] \p^a\nonumber\\
&-& \frac{1}{8}\int d^4x \left.\oo^3(y)f^2(y)\p \partial_y \p^a
\right\vert_{y=L_0}^{y=L_1}\nonumber\\
&+& \textrm{terms in $f(y)$ only}
}Fourier transforming in 4D  the pion field $\p^a(x,y) = \hat{\p}^a(q^2)
\s(q^2,y)$
the equation of motion and boundary condition is,
\eqs{
0&=&\left[ f^2(y)\oo^3(y) q^2+\partial_y \oo^3(y)f^2(y) \partial_y
\right] \s(q^2,y) \\
0&=& -\frac{1}{8}\left.\oo^3(y)f^2(y)\partial_y \s(q^2,y)
\right\vert_{y=L_1}.
}The resulting boundary theory is
\eqs{
&&\sum_a\int d^4q\hat{\p}^a(q) \S(q^2) \hat{\p}^a(-q) ,\\
\S(q^2)&=&\left. \frac{1}{8}\oo^3(y)f^2(y)\frac{\partial_y
\s(q^2,y)}{\s(q^2,y)}\right\vert_{y=L_0}.
}

For the simplest case it is instructive to demonstrate that the pion field
is a positive-norm state.  This will provide a starting point for the
discussion of more sophisticated models later in the paper.  The equation
of motion for $\s(q^2,y)$ can be solved and substituted into $\S$. 
Expanding at small $L_0$ and assuming $\D\ne 1$,
\eq{
\S(q^2) = \left[ \left(\frac{L_0}{L_1}\right)^{2\D-2}\!\!\!\!\! +\!
\frac{1}{4}+\!\mathcal{O}(L_0)\right]\frac{\U^2 L^3}{(\D-1)L_1^2}q^2 .
}There are two regimes to consider.  If $\D>1$ then the $L_0^{2\D-2}$ term
goes to zero as $L_0\rightarrow 0$, and the resulting co-efficient of the
$q^2$ term is positive definite implying that the theory is healthy.    If
$\D<1$ then the $L_0^{2\D-2}$ term in $\S$ diverges as $L_0\rightarrow 0$
and the $q^2$ co-efficient is negative indicating a negative-norm state. 
The theory is sick when $\D<1$.

If $\D=1$ then the equation of motion reduces to that of the vector field
eq.  (\ref{Veqnmotion}).  The solution for $\S$ becomes identical to the
unrenormalised expression for $\Pi_{V}$ in eq. (\ref{PiV}).  In the
expression for $\Pi_V$ is a log-divergence.  Effectively the $L_0^{2\D-2}$
 term  in $\S$ when $\D \ne 1$ `becomes a logarithmic divergence' when
$\D=1$.  The expression for $\S$ is proportional to $q^2\log L_0/L_1$.  As
$L_0 \rightarrow 0$ the state becomes non-normalisable and so it decouples
from the theory -- it is a free field.  Notice,  it is a positive-norm
state.

The interpretation associated to the three $\D$ regions of the theory is
the expected picture for a scalar field by considering 4D field theories. 
The bound on $\D$ from  unitarity considerations is that $\D > 1$.  This
is illustrated by  the negative sign found in $\S$ in the $\D<1$ regime. 
For $\D=1$ the scalar field has the na\" ive scaling dimension of a
classical field.  Such a theory is  \textit{trivial}.  Finally the case
that $\D >1$ is the situation one usually considers in a `healthy' theory.
 Such a result illustrates that although there is no obvious dual theory
to the 5D model we study the boundary theory does share the features of a
typical 4D theory relevant to phenomenology.

This section has examined the most well-studied case in the literature,
consisting of the action eq. (\ref{action}) with the background and VEV
profile in eq. (\ref{simpleVEV}).  In summary the result obtained is that
the following statements are all equivalent:  $\D>1$, the pion is
positively normalised or that $\hat{S}$ is positive.  Any one of these
three statements implies the remaining pair.

Under the assumption that $\D>1$, the pion field is a positive-norm state
and  $\hat{S}$ is positive.  With the basic case analysed we can begin
adding additional terms to the model.  In doing we ask whether given
$\D>1$ can a model be found where $\S'(q^2) >0 $ and $\hat{S}<0$.

\section{Boundary Actions for Scalars\label{scalars}}
Now consider eq. (\ref{action}) with the addition of new piece consisting
of a boundary Higgs term
\eqs{\label{DH}
S_{DH} &=& \int \sqrt{g}d^4xdy \bigg[\d(y-L_0)\l_{UV}+
\d(y-L_1)\l_{IR}\bigg]\nonumber\\&\times &g_{\m\n}\Tr (D^\m H)(D^\n
H)^\dagger
}so that the action is now $S_0+S_{DH}$.

When $\l_{IR}=0$, and in the unitary gauge, $S_{DH}$ does not change the
expression for $\hat{S}$ in terms of Lagrangian parameters.  Though it
will change the spectrum and so the numerical value of $\hat{S}$ when the
values of physical parameters are set.  The constraint from requiring that
the mass of the lowest lying axial state be positive is that
\eq{
\l_{UV}>-\frac{L_1}{\D-1} \left(\frac{ L_0}{L_1}\right)^{3-2\D}.
}If $\S(q^2)$ is computed in the case of a global symmetry, analogous to
the computation in Sec. \ref{elemcase} then a situation can occur where
$\S'(q^2)<0$ at $q^2=0$.  This imposes a constraint that for the pion to
be positively normalised
\eq{
\l_{UV} >-\half  \frac{L_1}{\D -1} \left(\frac{ L_0}{L_1}\right)^{3-2\D}.
}The pion constraint is more restrictive than that from coming from the
requirement of positive boson masses.  Notice this means that it is
possible to have a pion field that is a negative-norm state but not see
any pathologies in the spectrum or $\hat{S}$.  In Sec. \ref{effmetrics}
this result will re-occur.

Now consider the case that $\l_{UV}=0$ in eq. (\ref{DH}) and $\l_{IR}$ is
unspecified.  In this case one is changing the IR boundary condition to
\eq{
P(q^2,y) = \frac{1}{8} \l_{IR}(g_L^2+g_R^2) \oo^3(y) f^2(y)
}for the axial field.  Resolving the equations of motion in this case
leads to different expressions for $P_0$ and $P_1$ then by extension for
$\hat{S}$.  Full details of the computation are given in App. A.  Of
interest to this discussion is that provided $(g_L^2+g_R^2)L^2 \U^2 \ll 1$
\eqs{
\hat{S}&=&\frac{\e^2}{16L_1\D^2}g_L^2L^2\U^2\bigg[\D +1+\frac{2}{L_1}(\D
-1) \l_{IR} \bigg]
}By tuning $\l_{IR}$ one can produce $\hat{S}$ negative.   The bound to
produce $\hat{S}$ as positive is
\eq{
\l_{IR} > - \half L_1\frac{1+\D}{\D-1}
}Notice that this bound is negative.

Now study the pion field in the case of a global gauge symmetry and
$\l_{UV}=0, \, \l_{IR}$ unspecified.  By adding eq. (\ref{DH}) to the
action the IR boundary condition changes to
\eq{
\left[\partial_y \s(q^2,y)  - \l_{IR} q^2 \s(q^2,y)\right]_{y=L_1}=0.
}Re-solving the equations of motion and computing the new form of $\S$ gives
\eq{
\S(q^2) = \frac{L_1+2 (\D-1)\l_{IR}}{4L_1^3(\D-1)}\U^2 L^3 q^2 .
}In order that the pion be a properly normalised state
\eq{
\l_{IR}>-\half \frac{L_1}{\D-1}.
}This bound is more restrictive than the region of negative $\hat{S}$.

In the case that neither $\l_{IR}$ or $\l_{UV}$ is constrained one may try
to dial $\l_{IR}$ so that $\hat{S}$ is negative and then dial $\l_{UV}$ so
that the pion is nevertheless a positively normalised state.  However this
will not allow a healthy theory to be produced.  If one dials $\l_{IR}$ so
that $\hat{S}$ is negative then in order to obtain a positively normalised
pion one would require that
\eq{
\l_{UV} > -2L_1\frac{1+2(\D-1)\frac{\l_{IR}}{L_1}}{\D-1}\left(\frac{L_0}{L_1}\right)^{3-2\D} .
}

\begin{figure}
\includegraphics[width=0.5 \textwidth]{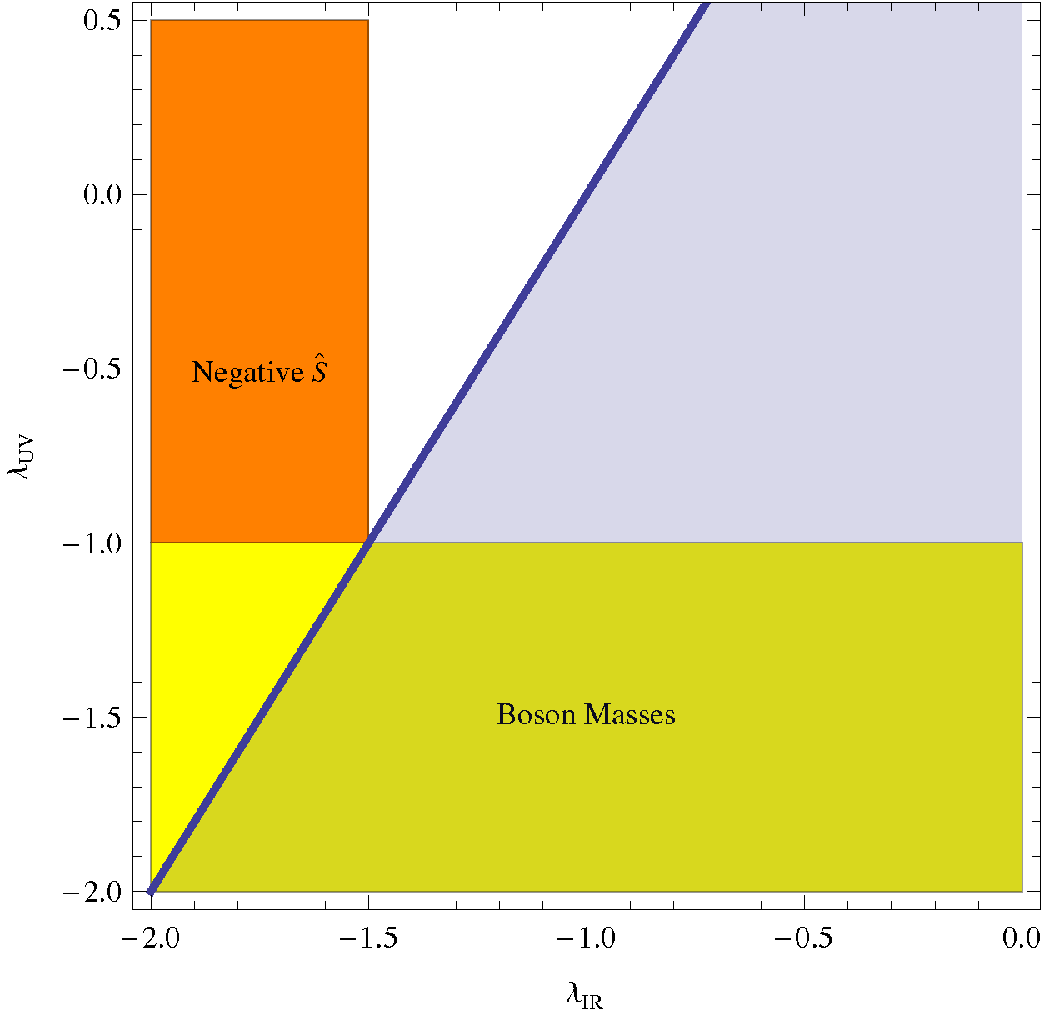}
\caption{\label{scalarplot}A plot of the $\l_{UV}$ with changing
$\l_{IR}$.  The blue region represents those points in parameter space
where the pion field is a positive-norm state.  The yellow region marks
where $\l_{UV}$ is so small as to produce unphysical gauge fields.  The
orange regions marks the points in parameter space where $\hat{S}$ is
negative.  The plot shows it is not possible to produce a healthy theory
with negative $\hat{S}$ by dialing scalar boundary term co-efficients.  For this graph $\l_{IR}$ has been normalised with a factor of $L_1$ to make it dimensionless and similarly a factor of $L_1 \left( \frac{L_0}{L_1}\right)^{2\D-3}$ has been included in $\l_{UV}$.}
\end{figure}

If $\l_{UV}$ is made as small as possible, to saturate the bound on
producing negative boson masses, then $\l_{IR}$ is still bounded to be above the point at which $\hat{S}$ becomes negative. This is illustrated in fig. (\ref{scalarplot}).

Therefore it is not possible to obtain $\hat{S}$ negative in a physically
acceptable theory by dialing a kinetic boundary term for the scalar
field.

\section{Boundary Actions for Vectors\label{boundaryvecs}}
Now consider eq. (\ref{action}) with the addition of a new term
that introduces an off-diagonal operator localised on the UV boundary. 
Such a term gives a
direct contribution to $\hat{S}$ and shifts the mass spectrum.

In doing this one must be careful not to break further symmetry.  Our
model contains a  bulk breaking of $SU(2)_L\times SU(2)_R \rightarrow
SU(2)_V$.  In adding an off-diagonal term to the UV boundary the mass
basis on the UV boundary may no longer be that of the axial and vector
fields -- the mass basis in the bulk.  If this were to happen then the
theory need not contain a spectrum of mass eigenstates with the quantum
numbers of the standard model.  For example adding the most obvious
operator, of the form $L^{\m\n}HR^{\r\s}H^\dagger$, to the UV boundary
would give a mass basis on the boundary that does not match the bulk mass
basis.

If $L^{\m\n}HR^{\r\s}H^\dagger$ is re-written as part of another term that
ensures any mass is given to the axial fields only (so that no further
symmetry breaking takes place) the symmetry breaking of the standard model
is ensured.  This can be done by using the operator
$[D^\m,D^\n]H([D^\r,D^\s]H)^\dagger$ .  Define
\eqs{\label{Shat}
S_{\hat{S}}&=&\int \sqrt{g}d^4xdy \d(y-L_0) \l_{\hat{S}}
g_{\m\n}g_{\r\s}\nonumber\\
&\times&\Tr ([D^\m,D^\r]H)([D^\n, D^\s]H)^\dagger
}so that the action we study is now $S_0+S_{\hat{S}}$.

To compute $\hat{S}$ one takes what could be termed the `bulk
contribution' originating from $S_0$ and
adds the direct contribution from $S_{\hat{S}}$. Combining the two
contributions
\eq{
\hat{S} =\e^2 \frac{1}{16} g_L^2 \frac{\D+1}{\D^2} \U^2L^2 + \half
\frac{g_L}{g_R} \l_{\hat{S}}\U^2 \left(\frac{L_0}{L_1} \right)^{2\D} \frac{L}{L_0} .
}Which implies that $\hat{S}$ is positive  whenever
\eq{
\l_{\hat{S}} > - \frac{1}{8}\frac{\D+1}{\D^2}\e^2 g_Lg_R L L_0\left(\frac{L_1}{L_0} \right)^{2\D}.
}
By introducing $\hat{S}$ the masses of the axial states are adjusted.  In
order that the gauge boson masses be positively normalised requires that,
\eq{
\l_{\hat{S}} >  -\frac{4}{\U^2} \frac{L}{L_0}\left( \frac{L_1}{L_0}\right)^{2\D}.
}This result requires a well-behaved expansion in $\U^2 (g_L^2+g_R^2)L^2$.
Notice that this bound does not directly constrain the sign of $\hat{S}$.  The two bounds on $\l_{\hat{S}}$ can be mutually satisfied to give a negative $\hat{S}$ in a healthy theory.

The mechanism that is employed can be understood in the following way. 
Using the action of $S+S_{\hat{S}}$, the parameter $\hat{S}$ is
schematically a bulk term plus a correction proportional to
$\l_{\hat{S}}$.  Meanwhile the $q^2$-coefficient of the axial vacuum
polarisation has a $\log$-divergence (which is renormalised by the
counterterm $Z$ into an $\e$-dependent term).  One can push $\l_{\hat{S}}$
very large and negative and still leave a properly normalised gauge field
because of the $\log$-divergence.  After renormalisation the divergence
`becomes the $\e$-term' in the vacuum polarisation.  It is then expected
that one may choose $\e$ in such a way as to properly normalise the gauge
field.  Meanwhile $\hat{S}$ can become negative as one dials
$\l_{\hat{S}}$ and $\e$.

Physically the scenario studied is of some higher theory that has a
contribution to $\hat{S}$.  This is parameterized by $\l_{\hat{S}}$ in the
effective description used here.  If this higher contribution were to
start out as negative then we have shown that the final value of $\hat{S}$
can remain negative, as one would expect.   It remains to be learnt
whether or not it is possible to write a higher-energy description (i.e.
moving beyond a simple 5D model) that can give fundamental contributions
to $\hat{S}$ that are negative.  Some distance towards this goal is
achieved in later sections.

\section{Recovering the Higgsless Models\label{limits}}
This section introduces an alternative logic to the existing literature on how one can understand the relationship between Higgsless models~\cite{Csaki:2003zu, Cacciapaglia:2004rb, Cacciapaglia:2004jz} and electroweak symmetry breaking models based around AdS/QCD~\cite{DaRold:2005zs, Erlich:2005qh, Hong:2006si, Fabbrichesi:2008ga, Piai:2007ys, Piai:2006hy, Round:2010kj}. 

To explain the idea this section introduces, consider the derivation of IR boundary conditions in this work as compared to the Higgsless models~\cite{Csaki:2003zu, Cacciapaglia:2004rb, Cacciapaglia:2004jz} or \cite{Hirn:2006nt}. 

In this work the IR boundary conditions of bulk fields are obtained by demanding that the variation of the action vanish on the IR boundary.  By placing terms localised on the IR boundary the conditions imposed on the bulk fields can be adjusted.  In this way one can arrange for whatever boundary conditions are appropriate.  While in a Higgsless models one simply enforces a boundary condition on the gauge fields -- the conditions are not the  result of some variation of the action vanishing.  

It is easy to write down an  IR boundary action for a Higgsless model so that the conditions arrived at by demanding the variation of the action vanish are the ones that were being enforced.  Going to the lengths of writing down a boundary action gives two benefits.  

Firstly, as has been extensively made use of in this work, one has a pion field that can be studied.  If the gauge field boundary conditions are not derived, then one does not know what the consistent set of conditions for the remaining (i.e. non-gauge) fields are.  As such, the bulk profiles are not fully known and the fields can not be studied.  The pion is an example of such a field.

Secondly the IR boundary action motivates one to think in an alternative way about AdS slice models that is more general than previous thinking.  In particular one is able to connect a large class of models as being special cases of a single model.  This is the idea that we wish to discuss further in this section.

To illustrate how models formulated on an AdS slice
can be connected, begin with the simpler task of producing a Higgsless model from eq. (\ref{action}).   There are two differences between the starting point of eq. (\ref{action}) and the goal of a Higgsless model.  Firstly there is no bulk breaking in a Higgsless model and secondly the IR boundary condition of the axial field must be changed.

Add an IR boundary term for the scalar field of the form
\eq{
\d(y-L_1)\l_{IR}|D_\m H|^2
}to the action eq.~(\ref{action}).  Our aim in adding this term is to
produce the boundary condition  $A_\m =0$ at the IR boundary; as is used
in the Higgsless model.  This boundary condition is achieved by taking $\l_{IR} \rightarrow \infty$.    This leaves the problem of the bulk scalar kinetic
term, which must be switched off -- but the IR boundary scalar kinetic
must be kept.

Therefore consider two limits.  Firstly the VEV profile $f$ is taken to
zero --- which can be achieved by dialing the overall numerical
co-efficient $\U$.  This removes the bulk breaking term.  However, to
prevent the boundary term $|D_\m H|^2$ from vanishing while taking $\U^2
\rightarrow 0$ keep $\z = \l_{IR}\U^2$ held fixed.  Then the limit to
produce a vanishing axial field on the IR boundary is taken by $\z
\rightarrow \infty$.  The resulting theory is of the same class of models 
as considered in the Higgsless scenario.

The procedure of limits promotes a way of thinking about the model in eq. (\ref{action}) and a Higgsless model.  A Higgsless model is a generic model of electroweak symmetry breaking formulated on a slice of AdS space where the IR bulk breaking is far stronger than any bulk breaking.  Conversely, the model of eq. (\ref{action}) is the case of a generic AdS model where bulk breaking is the most significant symmetry breaking.

An alternative to the limit procedure discussed here is to take $\D
\rightarrow \infty$ limit directly in eq. (\ref{action}).  This has the effect
of localising the bulk Higgs term to the IR wall.  In fact, rather than
thinking in terms of special cases where bulk breaking is small, as is
being promoted in this section, for the specific scenario of comparing the
model of eq. (\ref{action}) and the Higgsless case one can always take the limit
$\D \rightarrow \infty$ in any result derived from eq. (\ref{action}).   To see this is correct notice that the value $\D=1$ is always critical, above and below unity represent different physical scenarios.  Therefore increasing $\D$, when it is already larger than one , will not lead to new phenomena.  As a result one can take $\D \rightarrow \infty$ without encountering problems and indeed obtains the correct results.

Having seen an explicit example, the argument of this section can be stated
concisely.  Given a specific $\textrm{AdS}_5$ slice model, one can
consider it as a particular point in parameter space of the most general
model, that is one that includes all possible bulk and boundary terms.  In Sec. \ref{modbuild} it was discussed how to build the most general AdS slice model: one simply lists all possible operators that can live in the bulk and on the boundary then writes a model using operators up to some order in $q^2$.  By
considering various limits; for example where the bulk breaking is
sub-leading to the boundary breaking or  there is a significant
off-diagonal term present in the bulk, one can arrive at the specific
models in the literature.

As a further example the `effective metric' scenario of~\cite{Hirn:2006nt}
is produced as a special case of the most general AdS action
one could write down.  In this scenario a bulk term of the form
$L_{MN}HR^{MN}H^\dagger$ is needed in addition to the boundary scalar
kinetic term and the basic action eq. (\ref{action}).  As before (see Sec.
\ref{boundaryvecs}) an off-diagonal term is best included by adding a term
of the form
\eq{
\l_{\hat{S}}\Tr ([D_M,D_N]H)([D_P,D_Q]H)^\dagger
}into the bulk action\footnote{From the viewpoint of na\"ive dimensional analysis, one expects the parameter $\l_{\hat{S}}$ to be suppressed by three powers of the cut-off.  In this work we adopt a more basic approach of simply asking what is possible with such terms and so we allow $\l_{\hat{S}}$ to be arbitrary in magnitude.  The underlying mechanism that explains why such an operator might be larger than expected is beyond the scope of this work.}.  The procedure of limits then follows the previous
reasoning.  To remove the bulk kinetic term in eq. (\ref{action})  take
$\U^2\rightarrow 0$ while keeping $\z = \l_{IR} \U^2$ and $\l_{\hat{S}}
\U^2$ fixed.  This leaves a model with the correct field content and
Lagrangian terms of~\cite{Hirn:2006nt}.  In the limit that $\z \rightarrow
\infty$ the correct IR boundary condition is also obtained.  Therefore this scenario is also a particular case of the general AdS model one could begin with.

Finally, it is instructive to further study the procedure of limits
required to produce a Higgsless model and in particular the spectrum.  This section makes use of the ability to study the pion field when one uses boundary field variations to obtain boundary conditions.  

Once again consider the action required to produce a Higgsless model, this is eq. (\ref{action}) and an IR boundary scalar field.  In
the limit of $\U^2 \rightarrow 0$ with $\z = \U^2 \l_{IR}$ fixed the action becomes
\eqs{
S_{\U^2 \rightarrow 0}\!&=&\!\!\int \! \sqrt{g}d^4x dy \z' \d (y-L_1)
g_{\m\n} (D^\m e^{i\p})(D^\n e^{i\p})^\dagger  \nonumber\\
&-&
g_{MN}g_{PQ} \half \Tr\!\! \left(L^{MP}L^{NQ} + R^{MP}R^{NQ}\right),\\
\z' &=&\frac{1}{4}\left(\frac{y}{L_1}\right)^{2\D} \z .
}The spectrum of neutral states for this theory, in the limit that $\z
\rightarrow \infty$, is two light states, one vector and one axial.  These states correspond to the Standard
Model photon and $Z$.  Above that is a tower of masses approximately given
by the zeros of $J_\n (L_1 q)$ where $\n =1$ for the axial bosons and $\n
=0$ for the vector bosons.  For large values of $q^2$ the masses of the
axial and vector bosons will form two towers of states each tower having
the spacing between states of $ \p/L_1 $.  The two towers are
offset -- the lightest axial state that arises from the zeros of $J_0(L_1
q)$ is heavier by $\p/(2L_1)$ than the lightest vector state, that
originates from the zeros of $J_1(L_1 q)$.

Using $S_{\U^2 \rightarrow 0}$ the pion field can be studied by turning off the gauge couplings.  Solving the
equation of motion for $\s$ and applying the IR boundary condition $\s =0$
that results from taking $\z \rightarrow \infty$ gives
\eq{
\frac{\partial_y \s(q^2,L_1)}{\s(q^2,L_1)} = s_1 q^2 \rightarrow \infty
\textrm{ as } \z\rightarrow \infty
}where $s_1$ is the constant of integration.  Solving for $s_1$ requires
that $s_1 \rightarrow \infty$ which causes $\S(q^2)$ to diverge too.  The
pion profile has become a constant because of the limit $\U^2 \rightarrow
0$.   After normalising the pion kinetic term to unity the pion will
become a free field.  For this reason, that the pion is free in the limit
$\z \rightarrow \infty$, to study the `effective metric' scenario
of~\cite{Hirn:2006nt} one needs to use a large but finite $\z$ and
consider increasing $\z$ and the effects it has on the pion and $\hat{S}$.
 (For example, it may be that the pion is healthy at large and finite $\z$
in a region where $\hat{S}$ is negative and increasing $\z$ does not
change this.  Therefore one has evidence of a  region in parameter space
with negative $\hat{S}$ and a pion that appears to be healthy.)

\begin{figure}[hbp!]
\includegraphics[width = 0.5\textwidth]{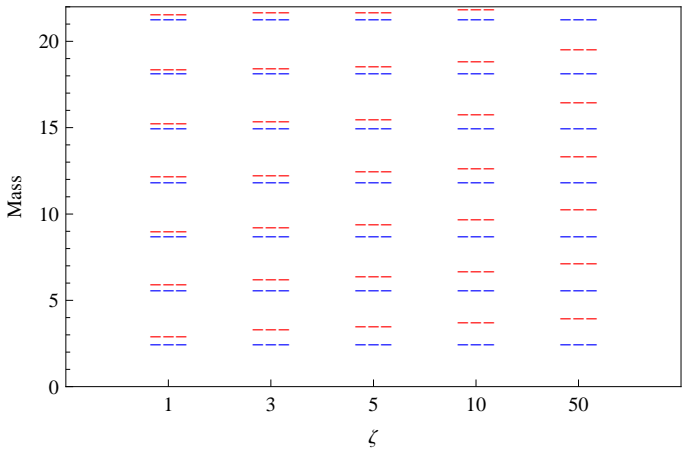}
\caption{\label{spectrum}The mass spectrum with increasing boundary Higgs
VEV $\z$.  The red lines represent the masses of the axial states and the
blue lines the masses of the vector states.  As $\z$ increases the axial and vector masses do not become approximately degenerate until higher mass.}
\end{figure}

Studying the pion field shows that by forcing $A_\m =0$ one produces a
theory with a free pion field.  As a result the symmetry breaking induced
by $H$ can not be restored at finite energy.  This is reflected by the
fact that the spectrum of the gauge bosons do not become degenerate at
high energies, instead the spacing is constant at all energies,
illustrated in  Fig. \ref{spectrum}.

\section{Effective Metrics\label{effmetrics}}
This section contains a  study of the effects of including a bulk
off-diagonal term.  To produce a scenario similar to that
of~\cite{Hirn:2006nt} we also include an IR boundary term.  Therefore we
add
\eqs{
S &=& \int \sqrt{g} d^4x dy g_{MN}g_{PQ}\Tr \bigg(
\nonumber\\ &-& 2\l_B \Tr ([D^M,D^P] H)([D^N,D^Q]H)^\dagger \bigg)\nonumber\\
&+&\d(y-L_1) \l_{IR}  g_{\m\n}\Tr  D^\m H(D^\n H)^\dagger .
}to eq. (\ref{action}) as the action we study in this section.  We
consider two cases, with and without the bulk Higgs kinetic term of eq.
(\ref{action}) switched on.

\subsection{The Limit of No Bulk Kinetic Term\label{effmet}}
Here the case that $\Tr |D_M H|^2$ is switched-off in the bulk is studied.
 Doing this keeps the number of free parameters in the survey under
control.

The logic of this section is that:  Given an action it is possible to
produce a negative $\hat{S}$ by dialing the co-efficient of a bulk
off-diagonal term (see~\cite{Hirn:2006nt} for examples).  However one must
check that the pion field is healthy, as has been focused on in this work.
 To control the study and not produce an overly complicated picture with
many free parameters we study $\hat{S}$ under the scenario that the bulk
breaking is weak.  In fact we will neglect the bulk term altogether.  When
examining the pion, the bulk kinetic term of the Higgs should \textit{not}
be turned off.   The idea that allows one to connect statements about
$\hat{S}$ in a theory with no bulk Higgs kinetic term and statements about
the pion is that the limit one takes to remove the bulk Higgs field can be
done smoothly and without affecting the sign of the pion kinetic term. 
This implies that one is analysing a scenario where the pion is healthy,
for example, and one restricts the analysis of $\hat{S}$ to  a
particular corner of parameter space where the bulk breaking is
sub-leading.

To switch the bulk kinetic term off, one takes the limit $\U^2\rightarrow
0$ whilst keeping $\z  = \l_{IR}\U^2$ and $\z_B(y)=\l_B(y)\U^2$ fixed (see
Sec. \ref{limits}).  In the axial and vector field basis the action
becomes
\eqs{
\int d^4x dy\!\!\!\!\! &&\!\!\!\!\!\frac{-1}{2} \ee_{MN}\ee_{PQ} \Tr[
(\oo(y)+\oo(y)\z_B(y))
V^{MP}V^{NQ}\nonumber\\&+&(\oo(y)-\oo(y)\z_B(y)A^{MP}A^{NQ}]
+\frac{1}{4}\z \left( \frac{y}{L_0}\right)^{2\D} \nonumber\\
&\times& \d(y-L_1)\ee_{\m\n}\Tr(D^\m e^{i \p} )(D^\n e^{-i\p})^\dagger .
}
Our study will be over a range of values for $\z$.  Of particular interest
is the limit $\z\rightarrow \infty$ where one obtains a similar scenario
to that found in~\cite{Hirn:2006nt}.   A large negative value of $\z$
leads to tachyonic axial masses.  Introduce `effective metrics' defined as
\eqs{
\oo_V(y) &=& \oo(y)\left[1+\z_B(y)\frac{y^{2\D}}{L_1^{2\D}}\right],\\
\oo_A(y) &=&
\oo(y)\left[1-\z_B(y)\frac{y^{2\D}}{L_1^{2\D}}\right].
}Using this notation the action becomes,
\eqs{
\int d^4x dy &-&\half \ee_{MN}\ee_{PQ} \Tr[ \oo_V(y)
V^{MP}V^{NQ}\nonumber\\&+&\oo_A(y)A^{MP}A^{NQ}]+\frac{1}{4}\z \left(
\frac{y}{L_0}\right)^{2\D} \nonumber\\
&\times& \d(y-L_1)\ee_{\m\n}\Tr(D^\m e^{i \p} )(D^\n e^{-i\p})^\dagger .
}From this, and comparing to the equations in Sec. \ref{elemcase} one can
read off the equations of motion for the axial  and vector fields.  In the
case of the pion, the equation of motion is the same as in previous
sections.  However as we will choose to specify the profiles $\oo_{A,V}$
one must find the metric by simultaneously solving the definitions of
$\oo_{A,V}$ for $\oo$.  Solutions to the axial and vector equations of
motion, in the limit of no Higgs kinetic term are discussed in Appendix
\ref{appvacpol}.

Discussing bulk off-diagonal terms analytically requires a survey of all
functions that tend to $1/y$ at small $y$.  Handling such a space of
functions and various integrals of them usually results in numerical
surveys.  Nevertheless one can make some progress if the metric profiles
$\oo_X(y)$ are assumed to be a monotonically decreasing function of $y$
(or more generally a decreasing function of sufficiently modest
variation).

Under this assumption, first examine the scenario $\z =0$ -- where there
is no electroweak symmetry breaking.  It will be shown that it is not
possible to obtain a negative contribution to $\hat{S}$ given that the
pion must be a positive-norm state.  Then the result is extended to $\z
\ne 0$.

Recall that in the case of $\z =0$ the pion vacuum polarisation at order
$q^2$ is (written in terms of $\oo_{A,V}$)
\eq{
\int_y^{L_1} (\oo_V(y)-\oo_A(y))(\oo_V(y)+\oo_A(y))^2dy
}For a positive-norm state we demand that this is positive definite which
can be ensured if $\oo_V > \oo_A \, \forall y$.  When applying this
constraint to $\hat{S}$ one can make progress by considering the cases of
$c_0$ and $\int dx/\oo_A(x)$ dominating the bracket $c_0+\int dx/\oo_A(x)$
(see eq. (\ref{nohiggsaxial}) ).  If $c_0$ dominates the bracket then at
leading order in $1/c_0$ the sign of $\hat{S}$ depends upon the difference
$\oo_V-\oo_A$ which is positive definite by choice.  Therefore only when the bracket in
$\Pi_A$ is dominated by $\int dx/\oo_A(x)$ (and $c_0$ is a correction) can $\hat{S}$ become negative.  In this case, the sign of $\hat{S}$ depends on whether
\eq{
\int \frac{dy}{\oo_A(y) }\int_y^{L_1}dx \oo_A(x)
> \int_y^{L_1} dx\left(\oo_A(x) \int \frac{dy}{\oo_A(y)}\right)
}with $\hat{S}$ positive if the inequality is satisfied.  Given a
decreasing function $\oo_A$, either monotonically or of sufficient rate,
this inequality is satisfied and $\hat{S}$ is positive.

When $\z \ne 0 $ one is changing the value of $c_0$.   However it has just been
argued that a negative $\hat{S}$ only occurs when corrections to $\hat{S}$ from $c_0$ are suppressed
relative to the contribution of metric factors themselves to $\hat{S}$(see previous paragraph).  In contrast, the line in $(\l_{IR}, \l_{UV})$ space that divides the two norm states of
the pion, positive and negative, is shifted as a leading order effect in
$c_0$.  As a result as one increases $c_0$ more and more of the $(\l_{IR}, \l_{UV})$ plane consists of negative-norm pion states.  Similarly there
are an increasing number of theories with negative $\hat{S}$ but the rate
at which the number of theories increases is lower that at which
theories are ruled out -- because the correction to $\hat{S}$ from changing $c_0$ is sub-leading.

In summary there are two cases to consider in understanding the expressions relating to $\hat{S}$.  In either case it is not possible to arrive at a situation where the pion is healthy and $\hat{S}$ is negative.

\begin{figure}
\includegraphics[width=0.5 \textwidth]{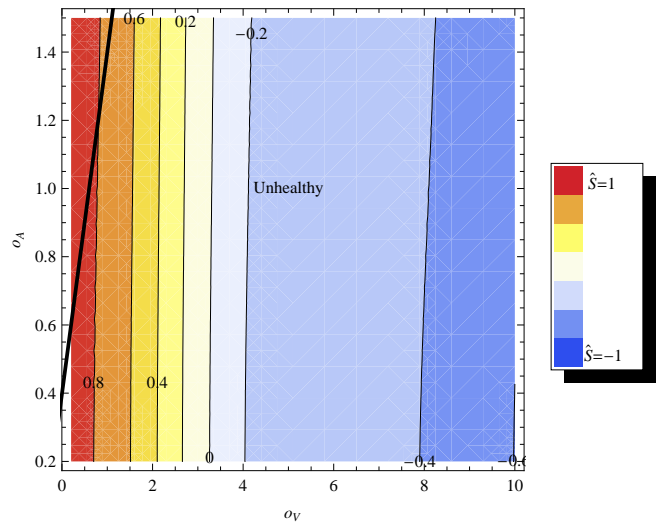}
\caption{\label{effective} A scan over the `effective metric' parameters
$o_{A,V}$ producing values of $\hat{S}$, indicated by colour.  A black
line marks the regions of space, marked `healthy' (`unhealthy'), where the
pion field is a positive (negative)-norm state.}
\end{figure}

By way of illustration a  survey of metrics with warp factors of the form
\eq{
\oo_X(y) = \frac{L}{y} \cos o_X y^2
}is given in fig. \ref{effective}.  The figure shows the $\hat{S}$
parameter as a function of $o_{A,V}$ with the constraint from requiring
that the pion be a positive-norm state added.    The value of the $o_A$ is
constrained to be positive and less than approximately 1.5 (beyond this
value the axial 2-point is not defined due to the specific metric profile choice).

The chosen form of metric pre-factor $1/y \cos o_X y^2$ can be thought of
as corresponding to $\D=2$ in the case of the earlier notation
eq.(\ref{simpleVEV}).  To see this consider the pion field with a non-zero
$f^2$, solve for $\z_B$
\eq{
\l_B f^2=\frac{\oo_V(y) - \oo_A(y) }{\oo_V(y)+\oo_A(y)}.
}If the expression is Taylor expanded for small $y$ then one has that
$f^2 \propto y^4$ which implies that in the UV one is examining a bulk
profile that is dual to condensate with anomalous dimension of
$\g_m=3-\D=1$, through the AdS/CFT dictionary.

\subsubsection{Infinite VEV Boundary Higgs}
Previous authors have considered the case that the boundary breaking term
is infinite.  As has been shown in sec. \ref{effmet} the theories considered that have
$\hat{S}<0$ contain a pion that is a negative-norm state.

This result has implications for the interpretation of~\cite{Hirn:2006nt}.  Using the results of this sec. \ref{effmet} one may
deduce that for the (bulk) pion field to be a positive-norm state; the
bulk spurion field used in \cite{Hirn:2006nt} is a different scalar to the
boundary symmetry breaking field.  This can be understood by considering a higher energy theory of which ~\cite{Hirn:2006nt} is the low energy effective description.  If the higher theory contains a contribution to $\hat{S}$  by some mechanism then this will show up in the low energy description as a fundamental contribution to $\hat{S}$.

\subsection{$W$ and $Y$ Parameters}

In principle, models based around eq. (\ref{action}) have uncontrolled
contributions to the $W$ and $Y$ electroweak precision
parameters~\cite{Round:2010kj}.  Having explored the possibility of
negative $\hat{S}$ it is interesting to consider how these models
contribute to the $q^4$-order precision parameters.  Restrictions on the
$W$ and $Y$ parameters from experiment imply an upper bound of $W,Y
\lesssim 10^{-3}$~\cite{Barbieri:2004qk}.   From na\" ive dimensional
analysis and applying the process of holographic
renormalisation~\cite{Round:2010kj} $W \sim  \mathcal{O}(\e^2 
m_W^2/M_\r^2)$ where $m_W$ is the $W$-mass and $M_\r$ is the $\r$ mass. 
The parameter $\e^2$ is the renormalisation parameter that scales as $N$. 
The mass of the $\r$ is limited by the $\hat{S}$ parameter.  The value of
$\e^2$ is constrained by requiring that the theory be at large-$N$
(implying large $\e^2$) and that $W$ and $Y$ be within experimental bounds
(implying a small or moderate $N$).  The relationship between the bulk and
boundary gauge coupling can then be used to understand whether some
consistent phenomenological scenario agrees with the assumptions of the
model.  That is, given that the precision parameters are of acceptable
values for some choice of $L_1$ and $\e^2$, is the bulk coupling
perturbative.  If a consistent picture is not available then one  can
increase the IR scale.  This lowers $\hat{S}$ and raise the $\r$ mass. 
The models discussed in this work have the potential to provide scenarios
where it is $W$ and $Y$ that are the most constraining of the precision
parameters, rather than $\hat{S}$.

Driving the IR cut-off to higher and higher values begins to imply fine
tuning.  For example it can be thought of as ``natural'' that the IR scale
is of order 1 TeV because it is related to the electroweak symmetry
breaking scale but increasing it greatly above such a value could be seen
as questionable.

An example of this behaviour is given by,
\eq{
\oo_X = \frac{L}{y} \cos o_X y^{\frac{5}{4}} .
}Using the reasoning that allowed us to connect an ``effective metric'' to
$\D=2$, this second profile, here, corresponds to $\D$ slightly greater
than one.  In the simple case of the VEV in eq.(\ref{simpleVEV}), it has
been shown~\cite{Round:2010kj} that choosing a small $\D$ constrains
$\e^2$ to be below one (after saturating the $\hat{S}$ bound) and thus
providing an example where $W$ is the most constraining parameter.

Whilst the example given is closely linked to existing models in the
literature, it serves to highlight the fact that in choosing arbitrary
profiles one will produce a large number of models where $W$ and $Y$ are
the precision parameters that most constrain the IR cut-off of the model.

\subsection{Inclusion of Bulk Higgs Kinetic Term}
In the most general case, that there is an off-diagonal bulk term, a
boundary breaking term and a bulk breaking term, again one may try to play
the two breaking schemes against one another and produce a negative
$\hat{S}$ and a healthy theory (similar to Sec. \ref{scalars}).  This
study  will be entirely numerical and have a  number of free parameters
that would be hard to control.  For this reason a detailed study of this
most general case is beyond the realm of the present work.

\section{Conclusions}
This work has investigated  the possibility of a negative $\hat{S}$
parameter within models of electroweak  symmetry breaking formulated on a
5D slice of AdS.  Attention has been given to ascertaining whether a
particular scenario that gives a negative $\hat{S}$ is acceptable on
physical grounds.

In the simplest AdS model, which has been well-studied in the literature
already, it was demonstrated that the pion field associated to the
symmetry breaking is a positive-norm state if and only if $\hat{S}$ is
positive.  In turn both of these statements are equivalent to the
statement that the scaling of the condensate, $\D$, is greater than one.

To consider  possibilities where  $\hat{S}$ is negative additional
operators were added to the UV and IR boundaries, then finally also to the
bulk.  This indeed allowed situations where $\hat{S}$ is negative.  Then,
within each scenario considered, the norm states of each field was
considered; which limited the parameter space.  This implied that it was
not possible to produce a negative $\hat{S}$ in a simple theory by
dialing Lagrangian parameters and requiring that the theory be healthy.

One particularly interesting scenario was the addition of a bulk
off-diagonal term.  In principle the most constraining of the precision
parameters of models which have off-diagonal contributions in the bulk can
be the  $W$ and $Y$ parameters.  This is in contrast to the usual picture
that $\hat{S}$ must be made to fall within bounds by dialing the IR
cut-off.  Instead one would need to dial the IR cut-off to make $W$ and
$Y$ fall within experimental bounds.  On the question of whether a
negative $\hat{S}$ could be produced in a theory containing healthy fields
it was found that this could only happen if one considered a situation
with a fundamental contribution to $\hat{S}$ that was negative.  This
would originate from a higher energy theory which the model studied in
this work would be a low energy effective theory of.

In conclusion this study has not found a consistent scenario that allows
for $\hat{S}$ to be negative and all states to have a  positive-norm,
without implying a fundamental and negative contribution to $\hat{S}$
originating from outside the electroweak symmetry breaking.  Our survey is
not totally complete, and it is possible that even higher order terms to
those considered, or an `effective metric' profile exists that provides a
healthy theory where $\hat{S}$ is negative.  We have demonstrated that
this would be an atypical theory and that caution should be exercised to
ensure that the theory does not contain a pathology of some kind.  In
addition we have illustrated that frequently apparently healthy scenarios
of negative $\hat{S}$ can be thought of as containing a fundamental
contribution to $\hat{S}$ that is large and negative.  In this light, it
seems unlikely that a generic scenario can occur where $\hat{S}$ is
negative due to the contribution of electroweak symmetry breaking.

\section{Acknowledgements}
This author wishes to thank M. Piai for useful discussions and interest in the project, plus comments on the manuscript.  This work was funded by the Science and Technology Funding Council (STFC) research grant ST/F00706X/1.

\bibliography{refs}
\appendix
\section{Solution to the Equations of Motion\label{appendix}}

Here a solution to the equations of motion for $P$, order by order in
$q^2$ are given.  Recall the equations of motion for $P$ are
\eqs{
\frac{1}{\oo}P_0' +\frac{1}{\oo^2}P_0^2
-\frac{1}{4}(g_L^2+g_R^2)\oo^2f^2&=&0,\\
\frac{1}{\oo}P_1' +2\frac{1}{\oo^2}P_0P_1 +1&=&0,\\
\frac{1}{\oo}P_2' +\frac{1}{\oo^2}\left( P_1^2 + 2 P_0P_2\right)&=&0.
}

\begin{widetext}When using the choices of eq. (\ref{simpleVEV}) and for
$\D \ne 1$
\eqs{
P_0(y) &=& \frac{ L}{ 2y^2}\frac{ I_{\frac{1}{\D}} \left(\frac{\a y^\D
}{\D}\right)  \G(1+1/\D)+ \a y^\D\left(\left[ I_{\frac{1}{\D}-1}
\left(\frac{\a y^\D }{\D}\right)  +  I_{\frac{1}{\D}+1} \left(\frac{\a
y^\D }{\D}\right)  \right] \G(1+1/\D)     +2 c_0 I_{1-\frac{1}{\D}}
\left(\frac{\a y^\D }{\D}\right) \G(1-1/\D)\right) }{I_{\frac{1}{\D}}
\left(\frac{\a y^\D }{\D}\right)\G(1+1/\D) +c_0I_{-\frac{1}{\D}}
\left(\frac{\a y^\D }{\D}\right)\G(1-1/\D)  }\\
P_1(y)&=& \exp \int_1^y -2\frac{x}{L}P_0(x)dx \left[ c_1 + \int_1^y
\frac{1}{x}\left(\exp \int_1^x 2\frac{z}{L}P_0(z)dz \right)dx\right]\\
\a^2 &=& \frac{1}{4}(g_L^2+g_R^2)\frac{L^2}{L_1^{2\D}}\U^2
}where the constants $c_{0,1}$ are found by imposing the boundary condition.

In order to substitute $P_0$ into the expression for $P_1$ first expand in
powers of $y$.
\eq{
P_0(y) =\half \frac{\a^2}{\D-1}y^{2\D-2}+\frac{1}{c_0} (-1)^\frac{1}{\D}
2^{1-\frac{2}{\D}}\left(\frac{\a}{\D}\right)^\frac{2}{\D} -
(-1)^\frac{1}{\D}c_0 2^{-1-\frac{2}{\D}} \a^2\D^2(\D+1)\left(\frac{\a}{\D}
\right)^\frac{2}{\D}y^{2\D}+\mathcal{O}(y^a;a>2\D)
}Schematically, as $y\rightarrow 0$, there is a term that diverges when
$\D<1$ giving an unbounded negative contribution to $P_0$, a constant term
and then higher terms that die away.  Substituting the expansion of $P_0$
into the expression for $P_1$
\eq{
P_1(y) = \log \frac{L_1}{y} + \half  L_1^2 \frac{1}{c_0} (-1)^\frac{1}{\D}
2^{1-\frac{2}{\D}}\left(\frac{\a}{\D}\right)^\frac{2}{\D} +
\frac{1}{4\D^2} L_1^{2\D} \frac{\a^2}{\D-1} + \mathcal{O}([L_1^\D
\a]^a;a>2)+\mathcal{O}(y)
}
Here, the constant of integration, $c_1$, has been solved for -- because
we are only interested in boundary conditions which lead to $P_1(L_1)=0$
in the main text.  The expansion in $L_1^\D \a$ is a more compact notation
for the constraint mentioned in the text that $\U^2 (g_L^2+g_R^2) L^2/4
\ll1$.
\end{widetext}

Alternatively one may wish to use a more general metric form.  In doing so
one will most likely obtain equations that require numerical solutions. 
An exception to this is when the bulk breaking term is switched off (or
considered weak compared to another source of symmetry breaking).  In the
scenario that it is the boundary conditions that break the electroweak
symmetry one can write down a remarkable set of equations.
\begin{widetext}
\eqs{
\Pi_V (q^2) &=&q^2\int_y^{L_1}\oo(y)dy+q^4 \int_y^{L_1}dx \frac{1}{\oo(x)}
\left[\int_x^{L_1} dz \oo(z)\right]^2+ \mathcal{O}(q^6)\\
\Pi_A(q^2) &=& \frac{1}{c_0+\int \frac{dy}{\oo(y)}}+ q^2\left(c_0 + \int
\frac{dy}{\oo(y)}\right)^{-2}\int_y^{L_1} dx \left[\oo(x) 
\left(c_0+\int\frac{dx}{\oo(x)} \right)^{2}\right]\nonumber\\
&+&q^4  \left(c_0 + \int \frac{dy}{\oo(y)}\right)^{-2}\int_y^{L_1} dx
\left[\frac{P_1^2(x)}{\oo(x)}  \left(c_0+\int\frac{dx}{\oo(x)}
\right)^{2}\right]\label{nohiggsaxial}
 +\mathcal{O}(q^6)
}with $P_1$ defined as $P_1 = \Pi_A '(0)$.
\end{widetext}
\section{Vacuum Polarisations\label{appvacpol}}

By solving the bulk equations of motion eq.(\ref{Veqnmotion}) for the
choice of eq. (\ref{simpleVEV}) and substituting into the definition of
the vacuum polarisation one obtains the unrenormalised expression for
$\Pi_V$.  Then the counterterm $Z$ is dialed to give
\eq{\label{PiV}
\Pi_{V}(q^2) = q^2 \left[ 1-\e^2 \left(\g_E+\log \half L_1\surd q^2
-\frac{\p}{2}\frac{Y_0(L_1\surd q^2)}{J_0(L_1\surd q^2)}  \right)\right]
}for $\g_E$ the Euler-Mascheroni number and $J, Y$ Bessel functions.  In
order to produce this expression a normalisation of $\Pi_V$ has been
chosen by choosing $\Pi_V'(q^2)=1$ at $q^2=0$.

The expression for $\Pi_V$ allows the spectrum to be read off.  There is a
massless photon and a tower of heavier states with the same quantum
numbers as the Standard Model photon.  The first such state will be
referred to as the $\r$ in analogy with QCD and has a mass $M_\r$
dependent upon $\e$.  The $\r$ mass for particular values of $\e$ can be
given as
\eq{M_\r \simeq \left\{\begin{array}{c l}
\frac{2.40}{L_1}&\textrm{for }\e \textrm{ small}\\
\frac{4.15}{L_1}&\textrm{for }\e = 1\\
\frac{4.69}{L_1}&\textrm{for }\e \textrm{ large}
\end{array}\right.
}

To obtain the axial vacuum polarisation, $\Pi_A$, the expressions for
$P_{0,1}$ are used.  The counterterm $Z$ removes the log-divergence
originating from $P_1$.  The result for $\D>1$ is
\begin{widetext}
\eq{
\Pi_A = \frac{\e^2}{L} \bigg[ \frac{1}{c_0} (-1)^\frac{1}{\D}
2^{1-\frac{2}{\D}}\left(\frac{\a}{\D}\right)^\frac{2}{\D} + q^2 \bigg(
\half  L_1^2 \frac{1}{c_0} (-1)^\frac{1}{\D}
2^{1-\frac{2}{\D}}\left(\frac{\a}{\D}\right)^\frac{2}{\D} + \frac{1}{4\D2}
L_1^{2\D} \frac{\a^2}{\D-1} \bigg)+\mathcal{O}(q^4)\bigg]
}
\end{widetext}
Assume that the first zero of the vacuum polarisation can be well
approximated by the ratio of the first two terms in the $q^2$ expansion of
$\Pi_A$.  Provided that $(g_L^2+g_R^2)L^2 \U^2 \ll 1$ and $\D>1$ the $Z$
mass is approximately
\eq{
m_Z^2\simeq  \frac{1}{8} \frac{1}{\D-1} \e^2 \U^2
(g_L^2+g_R^2)\frac{L^2}{L_1^2}.
}
\end{document}